\documentclass[preprint]{aastex631}

\begin{document}

\newcommand{\kms}{\ensuremath{\mathrm{km}\,\mathrm{s}^{-1}}}
\newcommand{\galunits}{\ensuremath{\mathrm{km}\,\mathrm{s}^{-1}\,\mathrm{kpc}^{-1}}}
\newcommand{\galacc}{\ensuremath{\mathrm{km}^{2}\,\mathrm{s}^{-2}\,\mathrm{kpc}^{-1}}}
\newcommand{\MLsun}{\ensuremath{\mathrm{M}_{\sun}/\mathrm{L}_{\sun}}}
\newcommand{\Lsun}{\ensuremath{\mathrm{L}_{\sun}}}
\newcommand{\Msun}{\ensuremath{\mathrm{M}_{\sun}}}
\newcommand{\Ha}{\ensuremath{\mathrm{H}\alpha}}
\newcommand{\SFR}{\ensuremath{\mathit{SFR}}}
\newcommand{\aveSFR}{\ensuremath{\langle \mathit{SFR} \rangle}}
\newcommand{\sfrate}{\ensuremath{\mathrm{M}_{\sun}\,\mathrm{yr}^{-1}}}
\newcommand{\Aunits}{\ensuremath{\mathrm{M}_{\sun}\,\mathrm{km}^{-4}\,\mathrm{s}^{4}}}
\newcommand{\surfdens}{\ensuremath{\mathrm{M}_{\sun}\,\mathrm{pc}^{-2}}}
\newcommand{\voldens}{\ensuremath{\mathrm{M}_{\sun}\,\mathrm{pc}^{-3}}}
\newcommand{\gevcc}{\ensuremath{\mathrm{GeV}\,\mathrm{cm}^{-3}}}
\newcommand{\etal}{et al.}
\newcommand{\LCDM}{$\Lambda$CDM}
\newcommand{\ML}{\ensuremath{\Upsilon_*}}
\newcommand{\Mst}{\ensuremath{M_*}}
\newcommand{\Mg}{\ensuremath{M_g}}
\newcommand{\Mb}{\ensuremath{M_b}}
\newcommand{\Mhalo}{\ensuremath{M_{\mathrm{halo}}}}
\newcommand{\Vhalo}{\ensuremath{V_{\mathrm{halo}}}}
\newcommand{\Vf}{\ensuremath{V_o}}
\newcommand{\sigstar}{\ensuremath{\sigma_*}}
\newcommand{\gobs}{\ensuremath{\mathrm{g}_{\mathrm{obs}}}}
\newcommand{\gtot}{\ensuremath{\mathrm{g}_{\mathrm{tot}}}}
\newcommand{\gbar}{\ensuremath{\mathrm{g}_{\mathrm{bar}}}}
\newcommand{\azero}{\ensuremath{\mathrm{g}_{\dagger}}}

\title{The Baryonic Tully-Fisher Relation in the Local Group and the \\ Equivalent Circular Velocity of Pressure Supported Dwarfs}

\author{Stacy S. McGaugh}
\affil{Department of Astronomy, Case Western Reserve University, 10900 Euclid Avenue, Cleveland, OH 44106, USA}

\author{Federico Lelli}
\affil{INAF -- Arcetri Astrophysical Observatory, Largo Enrico Fermi 5, I-50125, Firenze, Italy}

\author{James M. Schombert}
\affil{Institute for Fundamental Science, University of Oregon, Eugene, OR 97403, USA}

\author{Pengfei Li}
\affil{Department of Astronomy, Case Western Reserve University, 10900 Euclid Avenue, Cleveland, OH 44106, USA}

\author{Tiffany Visgaitis}
\affil{Department of Astronomy, Case Western Reserve University, 10900 Euclid Avenue, Cleveland, OH 44106, USA}

\author{Kaelee S. Parker}
\affil{Department of Astronomy, The University of Texas at Austin, 2515 Speedway, Stop C1400, Austin, TX 78712, USA} 

\author{Marcel S. Pawlowski}
\affil{Leibniz-Institut f\"ur Astrophysik Potsdam (AIP), An der Sternwarte 16, D-14482 Potsdam, Germany}

\begin{abstract}
We explore the Baryonic Tully-Fisher Relation in the Local Group. Rotationally supported Local Group galaxies adhere precisely 
to the relation defined by more distant galaxies. For pressure supported dwarf galaxies, we determine the scaling factor $\beta_c$ that relates their
observed velocity dispersion to the equivalent circular velocity of rotationally supported galaxies of the same mass such that $\Vf = \beta_c \sigstar$.
For a typical mass-to-light ratio $\ML = 2\;\MLsun$ in the $V$-band, we find that $\beta_c = 2$. More generally, $\log \beta_c = 0.25 \log \ML +0.226$.
This provides a common kinematic scale relating pressure and rotationally supported dwarf galaxies.
\end{abstract}


\section{Introduction}
\label{sec:intro}

Galaxies obey distinct kinematic scaling laws. 
Rotationally supported galaxies follow the \citet{TForig} relation that links luminosity with the outer circular velocity \Vf. 
Pressure supported systems follow the \citet{FJorig} relation that links luminosity to the stellar velocity dispersion \sigstar.
Though similar, the Tully-Fisher and Faber-Jackson relations are not identical for a number of reasons.
For one, the rotation speed $V$ of a disk and the velocity dispersion \sigstar\ of a spheroid are not identical measures. 
In the ideal case of isotropic orbits in a spherical system, the kinetic energy is split evenly among the three spatial dimensions 
and the equivalent circular speed of a test particle is $\sqrt{3} \sigstar$. 
{In a dynamically cold, rotationally supported disk with $\Vf/\sigstar \gg 1$, the measured rotation speed is already very close to
the circular speed of the gravitational potential, and can be corrected for modest non-cicrular motions as necessary.}
On top of this minimal difference {between pressure and rotationally supported systems}, 
the radii at which measurements are made varies widely. The velocity dispersion \sigstar\
of bright early type galaxies is typically measured in their high surface brightness centers where stars dominate the mass \citep{sauron}.
In rotationally supported galaxies, the approximately flat {circular} speed \Vf\ measured at the outermost radii {\citep[e.g.,][]{LelliTFscatter}}
provides the measure that minimizes the scatter in the Baryonic Tully-Fisher Relation \citep[BTFR;][]{LelliTF2019}. 
This typically occurs in the low acceleration regime \citep[$a < 3700\;\galacc$:][]{OneLaw}
where dark matter apparently dominates the mass budget. This difference makes it difficult to relate the star-dominated kinematics of the
Faber-Jackson relation to those of the BTFR. Nevertheless, there is a relation between \sigstar\ and \Vf\ among
early-type galaxies where both quantities can be measured \citep{Serra2016}, so there seems to be a connection.

In contrast to bright early type galaxies, the dwarf spheroidals of the Local Group reside predominantly in the low acceleration regime
of dark matter domination. This provides some prospect for relating pressure supported and rotationally supported systems on the same
characteristic velocity--mass relation. In this paper, we empirically identify the optimal value of $\beta_c$ in $\Vf = \beta_c \sigstar$ that
places Local Group dwarf spheroidals on the BTFR. This empirically motivated quantity is analogous to the flat 
portion\footnote{It is common in theoretical models to refer to dark matter halos by their circular speed $V_{200}$ at the virial radius or their
maximum circular velocity $V_{max}$ \citep{BBKARAA,WTARAA}. These quantities are not identical to \Vf.} of a rotation curve.

We construct the baryonic mass--{circular} speed relation for Local Group galaxies in \S \ref{sec:LGrot}, and 
check that the BTFR calibrated by external galaxies \citep{JSH0} applies to {rotationally supported} galaxies in the Local Group.
In \S \ref{sec:eqcirc} we identify a sample of dwarf Spheroidals for which we empirically measure the quantity $\beta_c$. 
We summarize our results in \S \ref{sec:conc}.

\section{The BTFR in the Local Group}
\label{sec:LGrot}

The quantities of luminosity and linewidth traditionally utilized for the Tully-Fisher relation
are proxies for more fundamental properties: the baryonic mass $\Mb = \Mst + \Mg$ and {outer circular} speed \Vf. 
The latter {quantities} define the BTFR \citep{M05}. {The scatter in the BTFR depends on how these quantities are measured.
Empirically, we have found that the scatter is minimized when near-infrared luminosities
are utilized to estimate stellar mass \citep{MS15} and when the outer velocity is measured from extended HI rotation curves.
See \citet{LelliTFscatter} for the algorithm by which the outer velocity is measured and \citet{LelliTF2019} for a comparison to other rotation speed measures.} 

The BTFR can be written as
\begin{equation}
\Vf = \left( 0.379\;\kms\,\Msun^{-1/4} \right) \Mb^{1/4}
\label{eq:BTFR}
\end{equation}
as calibrated with 50 galaxies with Cepheid or TRGB distances \citep[][]{JSH0}.
The dominant uncertainty is not from random errors in the fit but rather from systematics in the stellar mass estimates
and detailed corrections for metallicity and molecular gas \citep{metalsandmolecules}. 
We will determine $\beta_c$ by requiring that dwarfs adhere to equation \ref{eq:BTFR} in a statistical sense. 

The Tully-Fisher relation applies to rotationally supported galaxies. Prior to determining $\beta_c$ for pressure supported dwarfs in the
Local Group, we first check how the calibration of \citet{JSH0} compares with data for those members of the Local Group
that {have rotating gas disks}. This is a useful sanity check since the 50 calibrators of the BTFR are all external to the Local Group.

\subsection{Rotationally Supported Local Group Galaxies}

Rotationally supported Local Group galaxies are listed in Table \ref{tab:LGrot} in order of decreasing baryonic mass.
These are galaxies with the necessary data (e.g., atomic gas mass, some measure of the outer circular speed).
We restrict ourselves to traditional members of the Local Group \citep{mateo}, and do not include other nearby galaxies 
like those of the outlying NGC\,3109 association at $D\simeq1.3$ Mpc \citep[see][]{Pawlowski2014} 
or more distant ($\sim$2 Mpc) objects like NGC\,55, GR\,8, IC\,5152, and UGCA\,438. 
In particular, NGC\,3109 and NGC\,55 are in the SPARC sample \citep{SPARC} and have been 
used to set our baseline BTFR calibration \citep{JSH0}. The galaxies in Table \ref{tab:LGrot} 
have not been included in the BTFR calibration and are independent of it.

\begin{deluxetable}{lcccl}
\tablewidth{0pt}
\tablecaption{{Rotationally Supported Local Group Galaxies  \label{tab:LGrot}}}
\tablehead{%~\\
\colhead{Galaxy}  & \colhead{$\Mst$} & \colhead{$\Mg$} &  \colhead{$\Vf$} & \colhead{Ref.} 
\\ &   \multicolumn{2}{c}{($10^9\;\Msun$)}    & \colhead{($\kms$)}  & 
 }
\startdata
M31  &   135.\phn\phn\phn\phn &  \phn 5.46\phn &  $229.5\pm 2.2$      &      1 \\
MW   &   \phn 60.8\phn\phn\phn  & 12.2\phn\phn   & $197.9\pm 1.9$      &      2,3,4 \\
M33  &    \phn\phn  5.5\phn\phn\phn  & \phn 3.1\phn\phn    &  $118.0\pm 1.1$    &        5 \\
LMC &     \phn\phn 2.0\phn\phn\phn  & \phn 0.60\phn   &  $\phn 78.9\pm 7.5$      &      6,7,8 \\
SMC &	\phn\phn 0.31\phn\phn  & \phn 0.54\phn  &  $\phn56\pm 5$ & 6,7,9 \\
NGC 6822 &  \phn\phn\phn 0.234 \phd\phn   &  \phn 0.20\phn & $\phn55\pm3$ & 10 \\
%IC10    &    \phn\phn 0.1181  & \phn\phn 0.0271 & \dots    &       11 \\ %$\phd36.1\pm10$ for the desperate
WLM     &      \phn\phn\phn 0.0163\phn & \phn\phn0.077\phn  & $\phn38.7\pm3.4$   & 11,12 \\
%IC1613  &    \phn\phn 0.0288  & \phn\phn 0.0598 & \dots    &       11 \\ %$\phd16.9\pm12$ for the foolish
DDO 216  &      \phn\phn 0.0152 & \phn\phn\phn0.00816 & $\phn13.6\pm5.5$   &  11,12 \\      
DDO 210  &      \phn\phn\phn 0.00068 & \phn\phn\phn0.00274 & $\phn16.4\pm9.5$   & 11,12 \\
\enddata
\tablerefs{1.\ \citet{M31RC}. 2.\ \cite{LN2015}. 3.\ \citet{Olling} 4.\ \citet{Eilers2019}. 
5.\ \citet{Kam2017M33}. 6.\ \citet{skibba2012}. 7.\ \citet{BrunsLMC}. 8.\ \citet{LMCrot}. 
9.\ \citet{diT_SMC}. 10.\ \citet{weldrake}.  11.\ \citet{Zhang2012}. 12.\ \citet{iorio2017}.}
\end{deluxetable}

The literature contains many opinions about the relevant quantities for these well-{known} galaxies.
We utilize measurements that are compatible with the data for external galaxies yet independent of our own work. 
This {provides} a consistency check on the BTFR calibration of \citet{JSH0}. 
%The analysis in \S \ref{sec:LGbeta} utilizes only the latter.
We adopt a nominal error of 0.2 dex in mass to reflect the uncertainty in stellar masses stemming from the IMF and the foibles of SED fitting.

\subsubsection{Individual Galaxies}
\label{sec:indiegal}

Every galaxy is an individual with some peculiarities, so we give a brief description of each. 

\paragraph{M31} All the necessary information is provided by \citet{M31RC}. For the stellar mass, we adopt their stellar population (`SSP')
mass estimate as this is most consistent with the stellar population synthesis mass estimates that defines the mass scale of \citet{JSH0}.
\citet{M31RC} also discuss other stellar mass estimates that can dip slightly below $10^{11}\;\Msun$, illustrating the dominant systematic
uncertainty posed by stellar mass \citep[][]{BdJ,M05}.
The rotation speed is measured from HI observations reaching the nearly flat portion of the rotation curve beyond 100 arcminutes.
The rotation curve becomes dodgy beyond 130 arcminutes, so we neglect data beyond this radius which are well beyond the
levels reached for the external galaxies to which we compare. 

\begin{figure*}
\plotone{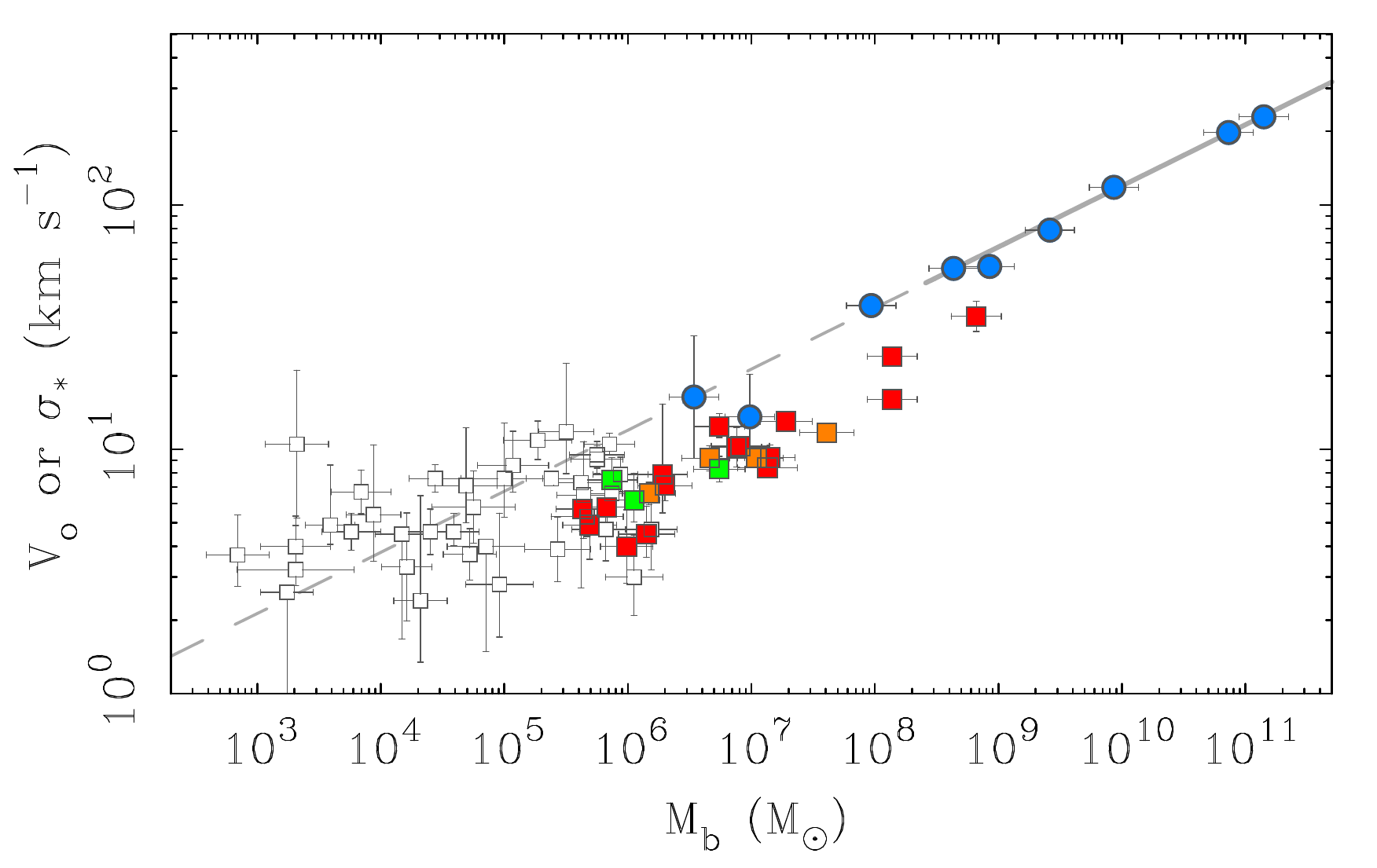}
\caption{The baryonic mass--{circular} velocity relation for Local Group galaxies. 
Rotationally supported galaxies with measured \Vf\ (circles, Table \ref{tab:LGrot})
are in good agreement with the BTFR calibrated independently with fifty galaxies external to the Local Group %that have Cepheid or TRGB distances
\citep[solid line; the dashed line is the extrapolation below the lowest mass calibrator]{JSH0}. 
Pressure supported dwarfs (squares) are plotted with their observed velocity dispersions \sigstar\ assuming $\ML = 2\;\MLsun$.
Filled squares are color coded by their proximity to M31 (red) or the Milky Way (orange) or neither (green).
Open squares are dwarfs whose velocity dispersions may not be reliable tracers of their equilibrium gravitational potential due to tidal effects (see text).
\label{fig:TF}}
\end{figure*}

\paragraph{Milky Way} We adopt the stellar population mass estimate of \citet{LN2015}. This {stellar mass estimate} 
is most comparable to the stellar masses employed to calibrate the BTFR, 
{and is nicely consistent with microlensing constraints on the IMF \citep{Wegg2017IMF}.
The adopted stellar mass} is heavier than some estimates \citep{BHGreview} but lighter than others \citep{PWmodelMW}, 
and is consistent with but independent of our own estimates \citep{M16,M19}. 
{It is bracketed by kinematic models of the Galactic dark matter halo for which $M_*$ can be either 
a bit higher or a bit lower depending on the choice of halo model \citep{SalucciMW}.}
The gas mass is obtained from integrating the surface density profile of \citet{Olling}
scaled to a modern size scale \citep{GRAVITY}. For the outer {circular} speed, we apply the method of \citet{LelliTFscatter} to the stellar rotation 
curve of \citet{Eilers2019}. The outer portion of the Milky Way rotation curve declines at the modest but perceptible rate of $-1.7\;\galunits$,
but a larger systematic is caused by the difference in the solar motion found by \citet{Eilers2019} and by \citet{M19}, with the latter being larger
by $\sim 4\;\kms$. We do not correct for this difference here to maintain independence. It is small and only worth noting because it exceeds the 
formal uncertainty, which, as always, should be taken with a grain of salt. This particular difference arises from a difference in the treatment of the
term for the gradient of the surface density in the Jeans equation. This difference reconciles an apparent discrepancy between the rotation curve 
obtained from stars and that from the terminal velocities of interstellar gas \citep[][]{M19}.

\begin{figure*}
\plotone{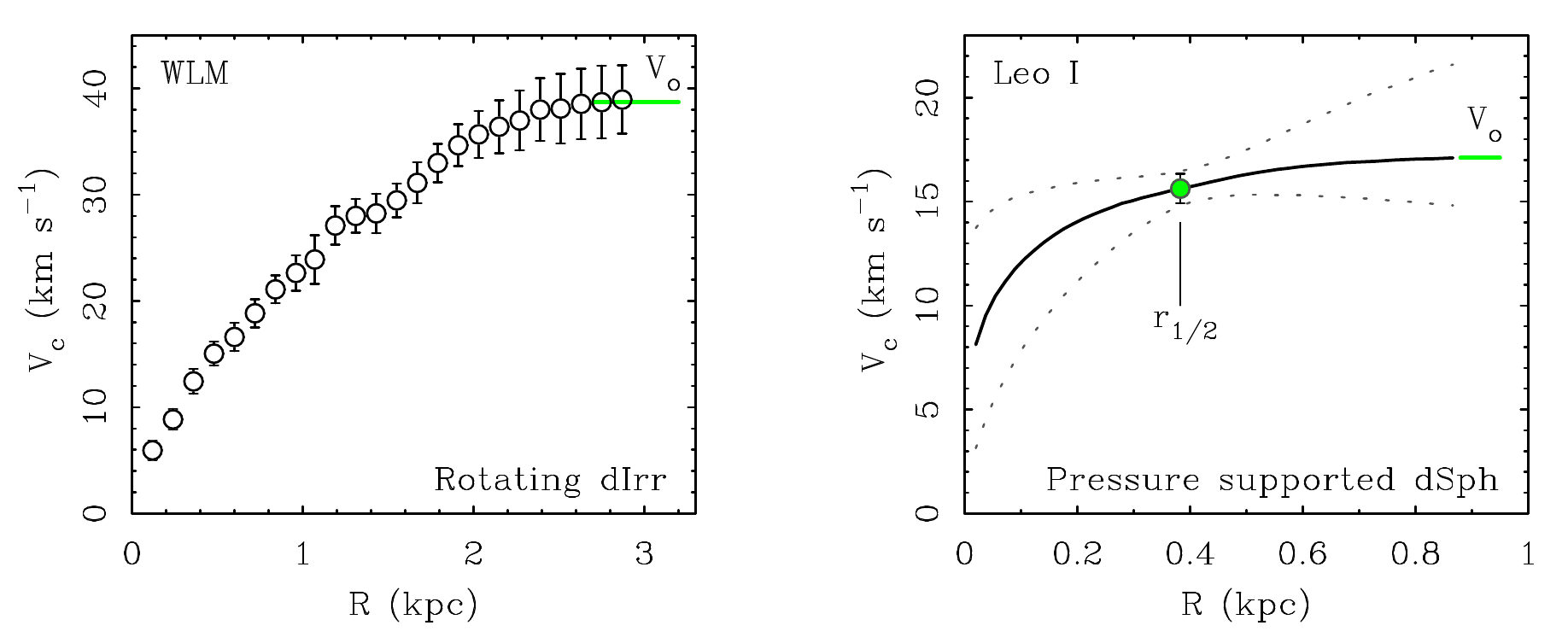}
\caption{The rotation curve of the gas rich Local Group dIrr WLM \citep[left]{iorio2017}
and the equivalent circular velocity curve of the pressure supported dSph Leo I (right).
The filled point represents the luminosity weighted circular speed $V_c = \sqrt{3} \sigstar$ at the 3D half light radius 
where variation due to anisotropy is minimized \citep{boom}.
The dotted lines illustrate how the uncertainty grows away from this point due to the compounding effects of anisotropy. 
The outer {circular} speed \Vf\ is marked for both. Note that $\Vf > \sqrt{3} \sigstar$ simply because of the shape of the circular velocity curve.
\label{fig:rot}}
\end{figure*}

\paragraph{M33} We adopt the measurements provided by \citet{Kam2017M33}. The stellar mass is based on the same near infrared
scale as our own stellar population estimates. We adjust the mass of atomic and molecular gas to account for helium and metals 
using the scaling relation of \citet{metalsandmolecules}. The {circular} speed is the average over the range
$9 < R < 16$ kpc where the {H\,I} rotation curve is {flat}. {The resulting \Vf\ is somewhat larger than that found by \citet{S1996}
{and \citet{Koch2018M33} but} slightly smaller than 
that\footnote{{\citet{SalucciM33} find $V_0 = 130.2 \pm 1.0\,\kms$. This quantity is similar but not identical to our \Vf. 
Their $V_0$ is a parameter of a function (their eq.\ 18) that attempts to fit the entire rotation curve while our \Vf\ only quantifies the amplitude
of the outer part of the rotation curve \citep{LelliTFscatter}. 
Using the same data \citep{Corbelli2014}, we obtain $\Vf = 119.6 \pm 1.8\,\kms$. 
The difference is entirely a matter of definition; compare the solid and dashed lines in Fig.\ 2 of \citet{SalucciM33}.}} of \citet{SalucciM33}.}

\paragraph{LMC} The LMC is clearly interacting with the Milky Way, so one may not expect it to retain equilibrium kinematics. Nevertheless,
it falls close to the BTFR \citep[for examples of other perturbed systems, see][]{verhTF}. We adopt the stellar mass estimate of \citet{skibba2012}
as being reasonably comparable to our own stellar mass scale, but smaller \citep{McC2012} and larger \citep{LMCreview} estimates can be found.
The same goes for the gas mass, the boundaries of which are challenging to demarcate given the Magellanic stream. We adopt the gas mass 
of \citet{BrunsLMC}, who take care to distinguish between gas in the LMC, SMC, and that in the Magellanic stream. 
In this and all subsequent cases, we correct the atomic gas mass for the hydrogen mass fraction and molecular gas as
described in \citet{metalsandmolecules}. For the {circular} speed, we adopt the measurement of \citet{LMCrot}
{Gaia proper motions}.

\paragraph{SMC} The dynamical status of the SMC is even more precarious than that of the LMC, and it shows in published velocity fields.
We again adopt the stellar mass estimate of \citet{skibba2012} and the gas mass of \citet{BrunsLMC}. For the rotation speed, we adopt 
the value reported by \citet{diT_SMC} {from H\,I}.

\paragraph{NGC 6822} All of the required information is provided by \citet{weldrake}. The stellar mass-to-light ratio that they adopt is far from consistent
with our own, so we adjust it to a $K$-band value of $\ML = 0.63\;\MLsun$ \citep{MS14}.

\paragraph{WLM} Data for the dIrr WLM are provided by \citet{iorio2017} who take their stellar masses from \citet{Zhang2012}. 
This is perhaps the lowest mass Local Group galaxy with a reliable measurement of \Vf.

\paragraph{DDO 210{, also known as the Aquarius dIrr}}  {All of the required data} are provided by \citet{iorio2017}. The kinematics have large uncertainties.
{The observed H\,I rotation velocity is extremely small ($\sim 5\;\kms$), comparable to the velocity dispersion of the gas ($\sim 6\;\kms$).
This leads to a large uncertainty in the asymmetric drift correction necessary to obtain the circular speed \Vf.} 

\paragraph{DDO 216{, also known as the Pegasus Dwarf}} Data are adopted from \citet{iorio2017} and are very uncertain.
Both DDO 216 and DDO 210 are transitional objects \citep{mateo} that may be 
subject to gas stripping: the data may not provide a reliable indicator of the circular speed of their equilibrium gravitational potentials.

There are other dwarf Irregular denizens of the Local Group, but the available kinematic data, if it exists at all, is of even lower quality than that of DDO 216.
For example, IC\,1613 and Sagittarius DIG do not have reliable estimates of \Vf\ because they are close to face-on.
Others have complex HI distributions that are manifestly out of dynamical equilibrium, e.g., IC\,10 \citep{IC10messy,IC10messier}. 
For further examples, see \citet{littlethings} and \citet{Oh2015}. It is challenging to obtain reliable tracers of the equilibrium gravitational potential
of very low mass galaxies, even those that are very nearby.
 
\subsection{Consistency Check}

The data for the Local Group rotators are shown in Fig.\ \ref{fig:TF} along with the calibration of \citet{JSH0}.
Agreement between these independent data is satisfactory. Indeed, the galaxies with the most reliable kinematics
--- M31, the Milky Way, M33, NGC 6822, and WLM --- 
adhere almost perfectly to the relation. It is hard to 
imagine\footnote{This agreement also makes it difficult to imagine a large systematic calibration error in the 
BTFR determination of $H_0 = 75.1 \pm 2.3\;\kms\,\mathrm{Mpc}^{-1}$ \citep{JSH0}.} better agreement.

%The BTFR is also in excellent agreement with the vast majority of low mass rotators in the field \citep{SHIELD,iorio2017} and holds as a continuous relation over five decades in mass down to the lowest mass rotator known, Leo P \citep[$\Mb \approx 10^6\;\Msun$][]{LeoPdisc,LeoProt,LeoPRhode,LeoPunQ}, 

\begin{figure*}
\plotone{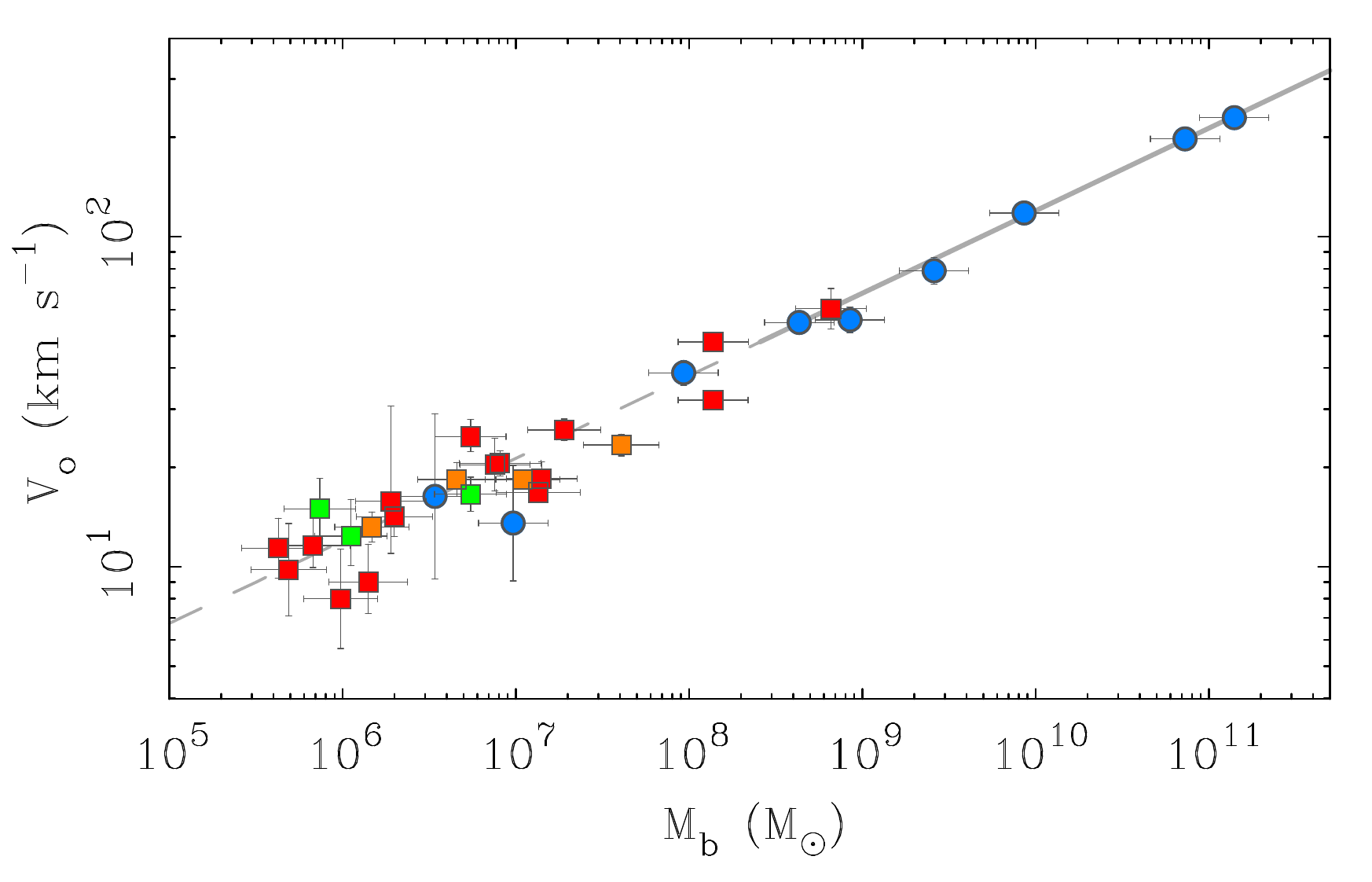}
\caption{The Baryonic Tully-Fisher relation of Local Group galaxies. Symbols have the same meaning as in Fig.\ \ref{fig:TF}. 
For the dwarf spheroidals, we plot $M_b = \ML L_V$ for $\ML = 2\;\MLsun$ and $\Vf = \beta_c \sigstar$ with $\beta_c = 2$.
The one exception is the most massive dwarf, NGC 205, for which $\beta_c = \sqrt{3}$.
\label{fig:LGTF}}
\end{figure*}

\section{The Equivalent Circular Velocity}
\label{sec:eqcirc}

In order to properly compare rotationally and pressure supported dwarfs, we would like to compare apples with apples. 
That the circular speed corresponding to an observed velocity dispersion is $V_c = \sqrt{3} \sigstar$ relies not only on the assumption of isotropy,
but also on an implicit assumption that both quantities are measured at the same radius. 
{This is not expected to be the case.
The outer orbital speed \Vf\ of rotationally supported galaxies is measured at the largest radii available from interferometric 21 cm 
observations \citep{LelliTFscatter} that tend to extend beyond the edge of the stellar disk.
In contrast, the velocity dispersions of dwarf galaxies in the Local Group are measured from spectra of individual stars.
A fair sampling of such tracers would place half of them inside the half-light radius, which is significantly less than the radii probed by \Vf.
We therefore do not expect} the common assumption of $\beta_c = \sqrt{3}$ to suffice to make an apples to apples comparison of \sigstar\ with \Vf.

In practice, {both} the number of stars observed in each dwarf (which can range from a {mere} few to thousands) 
and their locations within each dwarf vary widely {from case to case. 
Sometimes it is possible to obtain detailed velocity dispersion profiles $\sigstar(r)$ \citep[e.g.,][]{walker07}, 
and the considerations here are unnecessary. More commonly, however, 
one has perhaps a dozen stars that suffice to define a single velocity dispersion at whatever location the observed stars
happen to reside. This is taken to be characteristic of the global properties of the system, but} the effective 
radius of {such} measurements is not equivalent to the outer velocity measured in rotationally supported galaxies. 
So in addition to the issue of orbital isotropy, the value of $\beta_c$ also accounts for {differences in} the effective radius. 

{One convention is to} reference the velocity dispersions of dwarf spheroidals to their half-light radii, $\sigstar(r_{1/2})$ \citep{boom}. 
{This is equivalent to the luminosity-weighted velocity dispersion that one would obtain from the unresolved spectra of more distant galaxies.
By calibrating $\beta_c$ locally, we hope to extend its application to future discoveries of dwarfs beyond the Local Group.}

Figure \ref{fig:rot} shows the rotation curve of the Local Group dIrr WLM \citep{iorio2017} and the equivalent circular speed curve for the
dwarf spheroidal Leo I. The latter depends on the anisotropy of the orbits, which is generally unknown. 
One could view anisotropy as a contributor to the uncertainty in the circular speed that grows away from the half light radius \citep[Fig.\ \ref{fig:rot}]{boom}. 
Rather than try to estimate the likely range of anisotropy amongst dwarfs, we instead choose to ask the data.
The outer velocity is related to the measured velocity dispersion; what value of $\beta_c$ reconciles pressure supported dwarfs with the BTFR?

The assumption we make is that galaxies of the same baryonic mass have the same quasi-flat outer 
velocity\footnote{{See \citet{Serra2016} for a discussion of this point in high-mass early type galaxies.}}
irrespective of morphology. This is empirically motivated {\citep[e.g.,][]{URC2}, and the obvious assumption to make \citep[e.g.,][]{MMW98}. It} 
is reasonable in any theory. 
In \LCDM, dark matter halos of the same mass have the same structure \citep{NFW}. 
This is what is being probed so long as a galaxy is dark matter dominated. 
Similarly, a strict relation between velocity and mass is imposed by the force law in MOND \citep{milgrom83}
provided that the object is in the low acceleration regime. Being in the low acceleration regime is equivalent to being dark matter dominated in the 
conventional sense. {Not all galaxies are dark matter dominated \citep{URC2,MdB98a,StarkmanMaxDisk}, but} all of the dwarfs considered 
here\footnote{M32 is not considered here as its compact nature makes it star dominated and more akin to giant early type galaxies.} 
meet {the low acceleration} criterion with the exception of the brightest, NGC 205, which is just over the boundary \citep{OneLaw}.

\subsection{Tidal Demarcation}
\label{sec:rawTF}

The Local Group dwarfs divide into two families. The brighter dwarfs parallel the Tully-Fisher relation defined by rotationally supported galaxies 
(Fig.\ \ref{fig:TF}). The fainter dwarfs, most of which are the so-called ultrafaints \citep{simon2019}, break from the relation defined by the bright dwarfs, 
appearing to have little or no variation in velocity dispersion with luminosity.
The velocity dispersions of the ultrafaint dwarfs seem to saturate around $\sigstar \approx 6\;\kms$, albeit with a lot of scatter.

The break that is apparent in Fig.\ \ref{fig:TF} was noted by \citet{MWolf}. They provide a thorough discussion of the many reasons why it might arise.
One prominent possibility is tidal effects from the large, nearby hosts, Andromeda and the Milky Way. 
\citet{Bellazzini1996} suggested that the quantity $|M_V|+6.4 \log(D_{\mathrm{host}})$ was a good indicator of susceptibility to tidal effects.
This and other criteria were considered by \citet{MWolf}; all yield a similar result. For our purposes here, an effective
demarcation between the branches seen in Fig.\ \ref{fig:TF} is provided by 
\begin{equation}
|M_V|+6.4 \log(D_{\mathrm{host}}/\mathrm{kpc}) \lessgtr 23.  \label{eq:tides}
\end{equation}
Note that this is not simply a cut in luminosity, as there is considerable overlap among intermediate luminosity
dwarfs depending on the distance from their host. While tidal effects may not be the only reason
for the observed break, equation \ref{eq:tides} provides an effective criterion to distinguish between dwarfs 
that parallel the BTFR and those that do not. It is possible to determine a global value of $\beta_c$ that reconciles 
the parallel branch with the BTFR. It is manifestly not possible to find a single value of $\beta_c$ for those that do not.
The reasons why this might occur have been explored in depth by \citet{MWolf}. 
These are beyond the scope of this paper; here we simply 
exclude from further consideration dwarfs with $|M_V|+6.4 \log(D_{\mathrm{host}}) < 23$.

\subsection{Calibrating $\beta_c$ for Local Group Dwarf Spheroidals}
\label{sec:LGbeta}

To measure the globally optimal value of $\beta_c$ in $\Vf = \beta_c \sigstar$, we find the median value of $\beta_c$ that minimizes the difference between 
the BTFR and the dwarfs selected with the criterion specified by equation \ref{eq:tides}.  
Data for these dwarfs is given in order of decreasing baryonic mass in Table \ref{tab:LGdSph}. 
NGC 205 is excluded from the fit as it is not entirely in the low acceleration regime of dark matter domination. 
Intriguingly, it is already in good agreement with the BTFR for $\beta_c = \sqrt{3}$, as we might expect if the stars are important to the mass budget.

\begin{deluxetable}{lcccc}
\tablewidth{0pt}
\tablecaption{{Isolated Local Group Dwarfs  \label{tab:LGdSph}}}
\tablehead{%~\\
\colhead{Galaxy}  & \colhead{$\log(\Mb)$} &  \colhead{$\sigstar$} & \colhead{$\beta_c$} & \colhead{Host} \\
&   \colhead{($10^6\; \Msun$)}    & \colhead{(\kms)}  & &
 }
\startdata
NGC 205	&660.\phn\phn   & $35\pm5$ & \dots & M31 \\
NGC 185	&140.\phn\phn   & $24\pm1$ & $1.71\pm0.07$ & M31 \\
NGC 147	&140.\phn\phn   & $16\pm1$ & $2.57\pm0.16$ & M31 \\
Fornax	&\phn41.\phn\phn  & $11.7\pm0.9$ & $2.59\pm0.20$ & MW \\
And VII	&\phn19.\phn\phn   & $13.0\pm1.0$ & $1.93\pm0.15$ & M31 \\
And II	&\phn14.\phn\phn   & $9.25\pm1.1$ & $2.51\pm0.30$ & M31 \\
And XXXII	&\phn13.\phn\phn   & $8.4\pm0.6$ & $2.73\pm0.20$ & M31 \\
Leo I	&\phn11.\phn\phn   & $9.2\pm0.4$ & $2.37\pm0.10$ & MW \\
And XXXI	&\phn\phn8.1\phn   & $10.3\pm0.9$ & $1.97\pm0.17$ & M31 \\
And I	&\phn\phn7.6\phn   & $10.2\pm1.9$ & $1.95\pm0.36$ & M31 \\
Cetus	&\phn\phn5.5\phn   & $8.3\pm1.0$ & $2.21\pm0.27$ & \dots \\
And VI	&\phn\phn5.5\phn   & $12.4^{+1.5}_{-1.3}$ & $1.48^{+0.18}_{-0.16}$ & M31 \\
Sculptor	&\phn\phn4.6\phn   & $9.2\pm1.1$ & $1.91\pm0.23$ & MW \\
And XXIII	&\phn\phn2.0\phn   & $7.1\pm1.0$ & $2.00\pm0.28$ & M31 \\
Pisces 	&\phn\phn1.9\phn   & $7.9^{+5.3}_{-2.9}$ & $1.78^{+1.20}_{-0.65}$ & M31 \\
Leo II	&\phn\phn1.5\phn   & $6.6\pm0.7$ & $2.00\pm0.21$ & MW \\
And XXI	&\phn\phn1.4\phn   & $4.5^{+1.2}_{-1.0}$ & $2.90^{+0.77}_{-0.65}$ & M31 \\
Tucana	&\phn\phn1.1\phn   & $6.2^{+1.6}_{-1.3}$ & $1.99^{+0.51}_{-0.42}$ & \dots \\
And XV	&\phn\phn0.98   & $4.0\pm1.4$ & $2.98\pm1.04$ & M31 \\
Leo T	&\phn\phn0.74   & $7.5\pm1.6$ & $1.48\pm0.32$ & \dots \\
And XVI	&\phn\phn0.68   & $5.8^{+1.1}_{-0.9}$ & $1.87^{+0.36}_{-0.29}$ & M31 \\
And XXVIII	&\phn\phn0.49   & $4.9\pm1.6$ & $2.05\pm0.67$ & M31 \\
And XXIX	&\phn\phn0.43   & $5.7\pm1.2$ & $1.70\pm0.36$ & M31 \\
\enddata
\tablecomments{Mass and $\beta_c$ assume $\ML^V = 2\;\MLsun$.}
\end{deluxetable}

Our procedure averages over any anisotropies present in Local Group dwarfs.
It should provide a fair mapping of \sigstar\ to \Vf\ provided that there is no systematic alignment 
of orbital anisotropies in dwarfs along our line of sight \citep[which can happen in some models, e.g.,][]{Hammer18}.
Anisotropy in the orbits of stars in Local Group dwarfs is nevertheless an irreducible source of scatter in the BTFR.
For this reason alone, we expect more scatter for pressure supported systems than for rotationally supported systems.

Further scatter will be caused by variations in the stellar mass-to-light ratio from galaxy to galaxy.
In order to perform this exercise, we only need to know the mean stellar mass-to-light ratio 
for the dwarfs {so that we can} make an apples to apples comparison
with the BTFR (Fig.\ \ref{fig:LGTF}). We adopt a nominal $V$-band $\ML = 2\;\MLsun$ as a reference point,
motivated by {the color magnitude diagrams of resolved stellar populations} that suggest 
$\ML = 2.4\;\MLsun$ for Sculptor \citep{deboerSculptor} and $1.7\;\MLsun$ for Fornax \citep{deboerFornax,FornaxML}.
The resulting value of $\beta_c$ will be degenerate with the choice of \ML, so we derive an equation for the covariance that
enables the reader to choose whatever mass-to-light ratio seems best.

\begin{figure*}
\plotone{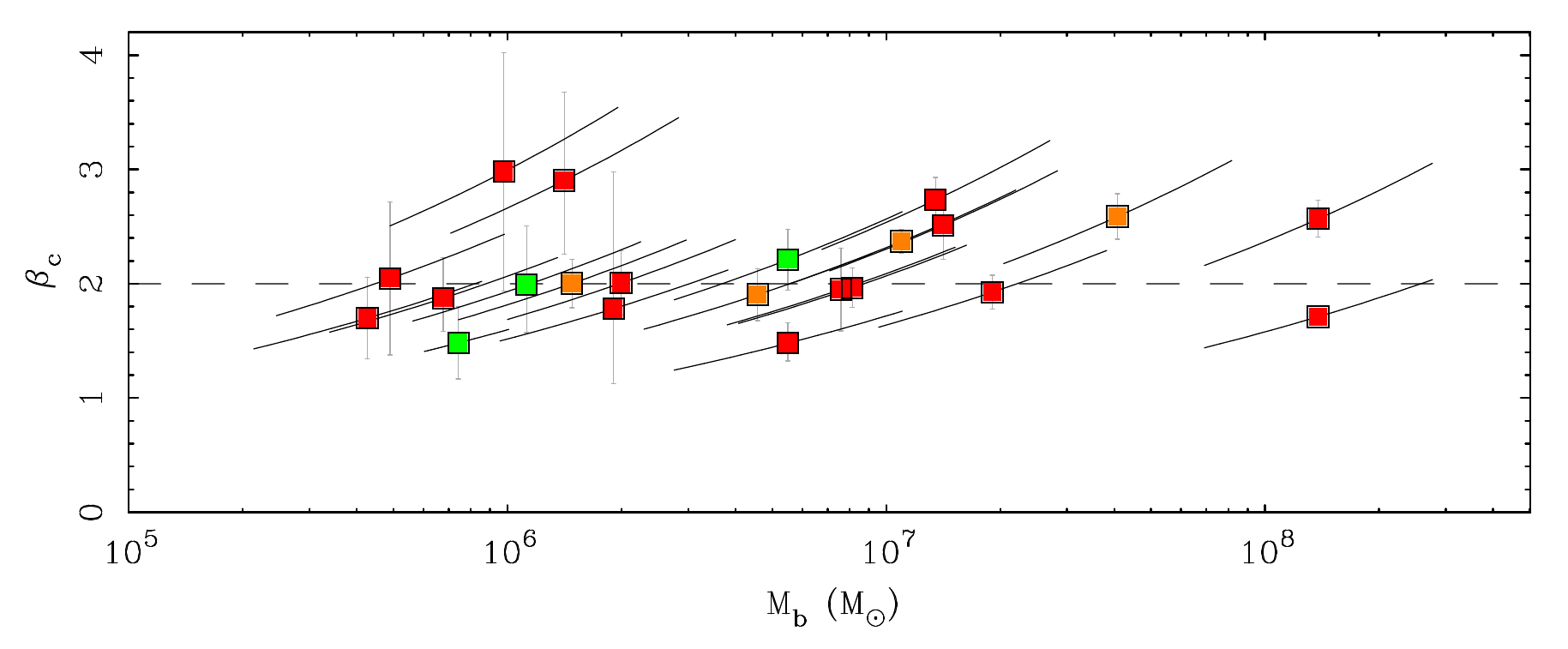}
\caption{The values of $\beta_c$ that place individual dwarfs precisely on the BTFR. 
Symbols have the same meaning as in Fig.\ \ref{fig:TF}. 
The central value of $\beta_c$ is shown as a point for $\ML = 2\;\MLsun$; the horizontal dashed line is the median $\beta_c = 2$.
Vertical error bars propagate the uncertainty in the velocity dispersion. Diagonal lines show the covariance with the mass-to-light ratio
illustrated by a factor of two variation from $\ML = 1$ (lower left) to $\ML = 4\;\MLsun$ (upper right).
%In the special case of Leo T, the range of variation is limited because much of its mass is in the form of gas, which is independent of the stellar \ML.
Variation in the stellar mass-to-light ratio from galaxy to galaxy contributes to the scatter, as does anisotropy.
\label{fig:DwBeta}}
\end{figure*}

The value $\beta_c$ that best matches each dSph to the BTFR is given in Table \ref{tab:LGdSph}.
The median $\beta_c = 2$ for  $\ML = 2\;\MLsun$ in the $V$-band. 
If instead we use the uncertainty-weighted biweight location, $\beta_c =1.97$. This difference is not significant.

The values of $\beta_c$ are plotted in Fig.\ \ref{fig:DwBeta}, which also shows its covariance with \ML.
This is given by
\begin{equation}
\log \beta_c = 0.25 \log \ML +0.226.
\label{eq:beta}
\end{equation}
Note that the slope of equation \ref{eq:beta} follows from that of the BTFR.
One may use this relation for any choice of mass-to-light ratio to find the value of $\beta_c$ that matches the measured velocity dispersions
of pressure supported dwarfs to the {circular} speeds of rotationally supported galaxies.

In general, we do not expect the flat portion of the circular velocity curve will be reached by the half light radius (Fig.\ \ref{fig:rot}).
If it did, we would expect $\beta_c = \sqrt{3}  = 1.73$. Instead, we typically expect $\beta_c > 1.73$. 
Treating this as a lower limit implies $\langle \ML \rangle > 1.12\;\MLsun$, which would indeed be rather
low for the $V$-band mass-to-light ratio of an old stellar population.

One can turn the question around, and attempt to infer variations in the mass-to-light ratios of dwarfs from their location above or below the
nominal relation. In doing so, the errors quickly explode due to the strong power-law relation between mass and {circular} speed.
The galaxy that deviates most clearly in this sense is Fornax, which has a low implied mass-to-light ratio. 
This is consistent with the observed presence of young stars in Fornax \citep{YoungFornax06,YoungFornax13,YoungFornax21}
and the mass-to-light ratio inferred from analyses of its color magnitude diagram \citep{FornaxML,deboerFornax}. 

The uncertainties are too large in most other cases to make any further inferences about variation in \ML. 
It is worth noting that the velocity dispersions of NGC 147 and NGC 185 (the highest mass galaxies in Fig.\ \ref{fig:DwBeta}) 
are clearly different even though they are indistinguishable in luminosity. This might be attributable to differences in the mass-to-light ratio, 
in their orbital anisotropy, or may simply be an indication of the limits of our method. 
It is also worth noting that these objects are on rather different orbits around M31 \citep{Sohn2020}, with NGC 147 showing distinct tidal
tails while NGC 185 does not \citep{Arias2016}.

{Another consideration is the distribution of tracer stars: different stellar populations may have
different radial distributions, which affects the term for the logarithmic density gradient in the Jeans equation. Indeed, \citet{slopesfrompops} 
describe how this effect can be used to leverage information about the mass profile of the dark matter halo.
Here, we are only making use of a single measured velocity dispersion from whatever tracers were available for observation,
in effect averaging over variations in the tracer distribution from galaxy to galaxy. These variations contribute to the scatter in $\beta_c$, 
which is modest.}

{Indeed,} there is no clear evidence of much intrinsic scatter in the remainder of the selected dwarf spheroidal data. 
This statement excludes the ultrafaint dwarfs
that were rejected by the tidal criterion of equation \ref{eq:tides}, as these clearly do deviate from the Tully-Fisher (Fig.\ \ref{fig:TF}). 
While tidal effects may not be the only explanation for the deviance of the ultrafaints \citep{mondultrafaints}, 
they seem like the most plausible candidate in the majority of cases \citep{MWolf}. 

Indeed, the situation for rotationally supported members of the Local Group (\S \ref{sec:indiegal})
drives home how challenging it can be to obtain robust kinematic probes of the gravitational potentials of low mass galaxies --- 
even those that are very near to us. It also makes one aware of the importance of tidal interactions: if relatively massive galaxies
like the LMC \citep{Besla2010} and Sgr dwarf \citep{Sgrdwarf} are subject to perturbation, what chance is there that ultrafaint dwarfs --- 
many of which are on plunging orbits that take them deep into the Milky Way potential \citep{orbits_simon} ---
remain unaffected by the same tidal forces? 

Finally, we note that in MOND we expect 
$\beta_c = (81/4)^{1/4} = 2.12$ for isotropic orbits in isolated dwarfs in the low acceleration regime \citep{milg7dw}. 
This corresponds to a mean $\langle \ML \rangle = 2.5\;\MLsun$, which seems plausible as an 
average mass-to-light ratio for old populations \citep{deboerSculptor}. 
This is only an approximate test of MOND, as the method employed here neglects the external field effect \citep[e.g.,][]{ChaeEFE}. 
The criterion imposed by equation \ref{eq:tides} will often but not always succeed in distinguishing dwarfs that should and should not
fall on the BTFR \citep[see discussion in][]{MWolf}. This is an important distinction in MOND, which is the only theory that has demonstrated the
ability to predict velocity dispersions in advance of their observation \citep{MM13a,MM13b,Pawlowski2014,Crater2pred,DF2mond}.

\section{Summary}
\label{sec:conc}

We have investigated the kinematics of pressure supported and rotationally supported galaxies in the Local Group.
We confirm that {rotationally supported} Local Group galaxies are in excellent agreement with the Baryonic Tully-Fisher Relation
calibrated with external galaxies. We further find that the velocity dispersions of pressure supported dwarf spheroidals can be related to the 
outer, quasi-flat {circular} speeds of rotationally supported galaxies through
$\Vf = \beta_c \sigstar$ with $\beta_c = 2$ for  $\ML = 2\;\MLsun$ in the $V$-band. 
We provide a more general formula for other choices of the stellar mass-to-light ratio (equation \ref{eq:beta}).
The median $\beta_c > \sqrt{3}$ likely indicates that the radius where the circular speed flattens out is greater than
the radius were \sigstar\ is typically measured. Correlated anisotropy along the line of sight could conceivably have the same effect.
Either way, our findings provide a unifying scale with which to discuss systems of differing morphology.

{These results are an indication of the tension between galaxy dynamics and cosmologically motivated collisionless dark matter.
The quasi-flat rotation speed \Vf\ occurs in the low acceleration regime of dark matter domination, and would seem to be a property of the dark matter
halo. This does not sit comfortably with the observational fact that \Vf\ does not depend on the dark matter fraction. It scales strictly with the baryonic
mass (Fig.\ \ref{fig:LGTF}) irrespective of whether a galaxy is dark matter dominated or not. 
The baryonic mass and the details of its distribution are far more strongly coupled to the dynamics of galaxies \citep[e.g.,][]{URC2,M20} 
than was anticipated by natural models involving halos of cold dark matter \citep[e.g.,][]{MMW98,MdB98a}.
More recent, more complicated models do not provide a satisfactory explanation for this simple phenomenology \citep{M21},
which remains poorly understood.}

%More generally, galaxy dynamics are entirely predictable from knowledge  of the baryon distribution alone \citep{URC2,M20}, even in cases where dark matter apparently dominates. This makes little sense in terms of non-baryonic cold dark matter that interacts only weakly with baryons and has a very different phase space distribution. There is no good reason why the dominant dark mass should be specified by that of the luminous baryons, yet that is what happens.}

\begin{acknowledgements} 
{We thank the referee for a number of helpful suggestions.} 
SSM thanks Joe Wolf for suggesting that we consider the calculated shapes of the effective circular velocity curves of pressure supported dwarfs
in the same way as the rotation curves of disk galaxies. 
The work of SSM, JMS, PL, and TV was supported in part by NASA ADAP grant 80NSSC19k0570 and NSF PHY-1911909.
MSP was supported by Leibniz-Junior Research Group grant J94/2020 via the Leibniz Competition, 
and a Klaus Tschira Boost Fund provided by the Klaus Tschira Stiftung and the German Scholars Organization.
\end{acknowledgements}

%\bibliography{LGTF_paper}
%\bibliographystyle{aasjournal}

\end{document}